\documentclass{goose-article}

\hypersetup{pdfauthor={T.W.J. de Geus}}
\hypersetup{pdftitle={An analysis of the pile-up of infinite periodic walls of edge dislocations}}

\header{%
  T.W.J.~de~Geus, R.H.J.~Peerlings, C.B.~Hirschberger\\%
  Mechanics Research Communications, 2013, 54:7-13, %
  \doi{10.1016/j.mechrescom.2013.08.010}%
}

\title{%
  An analysis of the pile-up of\\ infinite periodic walls of edge dislocations
}
\author{T.W.J.~de~Geus}
\author{R.H.J.~Peerlings$^*$}
\author{C.B.~Hirschberger}
\affil{
  Eindhoven University of Technology, Department of Mechanical Engineering,\nl
  P.O.~Box~513, 5600~MB~Eindhoven, The~Netherlands
}
\contact{%
  $^*$Corresponding author:
  \href{mailto:r.h.j.peerlings@tue.nl}{r.h.j.peerlings@tue.nl} %
}

\begin{document}

\maketitle
\sloppy

\begin{abstract}
We analyse the equilibrium pile-up configurations of infinite periodic walls of edge dislocations which are forced against an impenetrable obstacle by a constant applied shear stress. Numerically generated density distributions exhibit two distinct regions, for each of which we provide an interpretation and an analytical prediction. Near the obstacle, the influence of neighbouring slip planes may be neglected and the classical solution for a single slip plane applies. At a larger distance a linear decay is obtained. The characteristic length scales of the two parts of the pile-up are shown to depend differently on the parameters of the problem.
\end{abstract}


\section{Introduction}

Grain boundaries, second-phase particles and other heterogeneities in the microstructure of polycrystalline materials have a pronounced effect on the material's overall inelastic response. Their presence impedes the glide of dislocations and thus allows a smaller amount of local plastic deformation at a certain applied stress. Heterogeneities in the dislocation density which thus arise are the cause of so-called size effects, i.e.\ a dependence of the macroscopically measured mechanical response on the spatial scale (size) of the microstructure. A well-known example is the Hall--Petch effect \citep{Hall1951,Petch1953} of the average grain size on the yield strength of a polycrystal.

These observations, among others, inspired researchers from the mid-20th century onwards to theoretically study the pile-up of dislocations against impenetrable obstacles, see e.g.\ References~\citep{Eshelby1951,Leibfried1951,Head1955,Chou1961,Chou1967,Pande1970}. Due to the discrete nature of dislocations and the stress fields emitted by them, individual dislocations of the same sign (or orientation and Burgers vector) generally repel each other. This implies that, as some of them get stuck at an obstacle, those that follow remain at a finite distance from the first one and from each other. As result, a boundary layer is formed along the obstacle, with an increased dislocation density, which however decays with increasing distance from the obstacle -- a pile-up.

Many of the existing studies of pile-ups aim at characterising, or predicting, the dislocation density profile leading up to the obstacle. The earliest studies consider a linear array of edge or screw dislocations on a single glide plane. For this case, first studied in a discrete setting in the classical papers by \citet{Eshelby1951}, \citet{Leibfried1951} and \citet{Head1955} established a continuous solution which essentially shows a $1/\sqrt{x}$ dependence of the dislocation density on the distance $x$ to the obstacle. Subsequent studies of the interaction between linear pile-ups on different glide planes, or in fact on a family of glide planes, have shown that such interactions may significantly influence the density profile -- see e.g.\ \citep{Head1955,Chou1961,Chou1967,Louat1963,Pande1970}. In particular, \citet{Louat1963} established an analytical density distribution for an infinite stack of linear pile-ups of screw dislocations. This distribution differs significantly from the classical $1/\sqrt{x}$ decay and shows a dependence on the spacing of the glide planes on which the pile-ups live.

\citet{Roy2008} performed a numerical study of infinite walls of screw and edge dislocations. For this study, infinite dislocation walls piling-up against an obstacle perpendicular to their glide planes were selected because this problem has short-range stresses only, and is thus ideally suited to examine their effect. For screw dislocations, the numerically computed dislocation density profiles corresponded well with the analytical solution by \citet{Louat1963}. For edge dislocations, no closed-form solution appeared to be available in the literature. In the numerical simulations, the classical $1/\sqrt{x}$ decay of the dislocation density was observed close to the obstacle, which at somewhat larger distances transitioned into another, unknown dependence.

More recently, \citet{Hall2011a} has established, by a rigorous
discrete-to-continuum transition, that the density profile at some distance
from the obstacle becomes linear. Numerical solutions of the discrete problem
are shown to largely follow this linear dependence, apart from boundary
layers at the head and tail of the pile-up. Hall also discusses the physical
relevance of the assumption that the walls are periodic and concludes that
walls are likely to emerge on non-periodic (active) slip planes as well, but
the interactions between such non-periodic walls may be quite different from
the periodic case.

Related studies of parallel pile-ups have been done by \citet{Baskaran2010} and \citet{Schouwenaars2010}. In particular, \citet{Baskaran2010} studied a double-ended pile-up problem in which the slip planes are oriented at an arbitrary angle with respect to the obstacles. For angles other than 90 degrees, long-range stress fields exist that are shown to be dominant in the formation of the pile-up. In the ``degenerate'' case of exactly 90 degrees no such long-range stresses exist and the approach followed cannot be used. \citet{Schouwenaars2010} examined the influence of various idealisations -- in particular of the assumption of infinite dislocation walls and infinite dislocation lines. They correctly argue that finite walls of dislocations have long-range stress fields and that these may overwhelm the short-range stresses emitted by the individual dislocations -- see e.g.\ also \citep{Lubarda1996}.

In this paper we return to the case of infinite walls of edge dislocations piling up against a parallel obstacle considered also by \citet{Roy2008} and by \citet{Hall2011a}. We would like to emphasise that this case is highly idealised in many respects. Dislocation structures encountered in real materials obviously are unlikely to be perfectly periodic and neither are they infinite. But perhaps more importantly, in real materials multiple slip systems are available and additional mechanisms such as cross-slip, climb, etc.\ may become active. We nevertheless believe the idealised case of periodic single slip is worth studying more closely for three reasons: (i) it allows us to study the effect of short-range stresses in dislocation interaction in a clear, transparent setting, where it is not cluttered by other effects (see also \citep{Roy2008}); (ii) in particular, it allows us to study the influence of mutual interaction between different glide planes, depending on their spacing; (iii) it emphasises once more the importance of accounting properly for the discreteness of dislocations and their interactions \citep{Roy2008,Hall2011a}.

The purpose of this paper is threefold. Firstly, we confirm the linear density profile predicted by \citet{Hall2011a} via an alternative, more heuristic route. Secondly, we show that near the obstacle this profile transitions into the $1/\sqrt{x}$ dependence observed by \citet{Roy2008} and that this ``boundary layer'' (in the terminology of \citet{Hall2011a}) may thus also be described by a continuous density. And thirdly, we show that depending on the parameters of the problem, on of the two regimes may be dominant, or both occur at the same time. In the latter case, a fairly good prediction of the entire pile-up, including the transition point between the two regions, is obtained by combining the two analytical expressions.

The remainder of this paper is organised as follows. Section~\ref{sec:problem}
defines the discrete dislocation pile-up problem which we study
here and presents the numerical solutions which we use as a reference
throughout the paper. The two regions which can be distinguished in the
numerical data are analysed individually in Section~\ref{sec:near} and
Section~\ref{sec:remote}, whereas the transition between them is discussed
in Section~\ref{sec:transition}. We close with a brief summary of conclusions
in Section~\ref{sec:conclusion}.

\section{Numerical solution of the discrete pile-up problem}
\label{sec:problem}

\begin{figure}[htp]
  \centering
  \includegraphics[width=110mm]{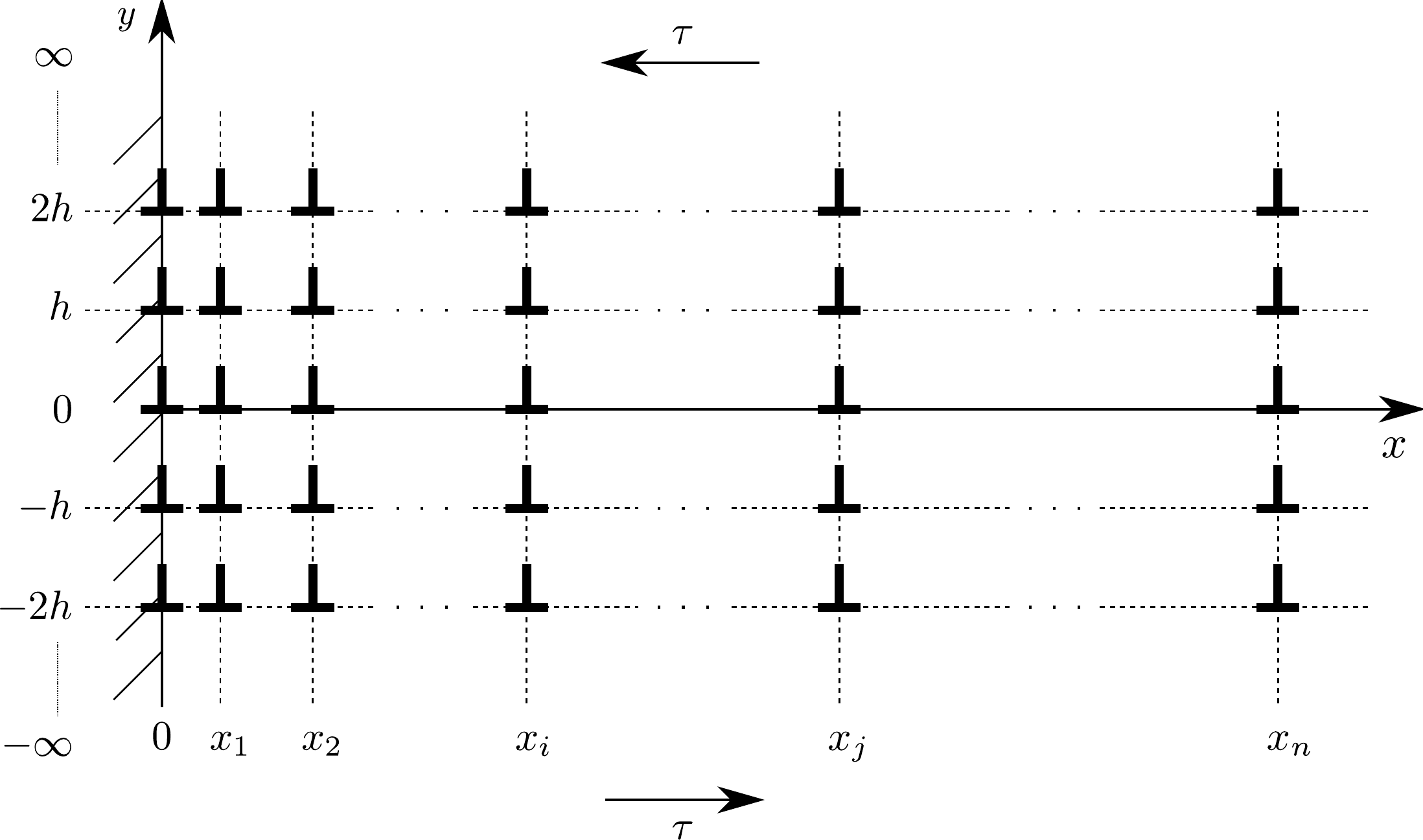}
  \caption{Pile-up of vertical walls of edge dislocations in an infinite elastic
  medium. The dislocation wall at $x=0$ is immobile.}
  \label{fig:geometry}
\end{figure}

We consider the problem of a pile-up of edge dislocations against an impenetrable obstacle, as sketched in Figure~\ref{fig:geometry}. The dislocations, denoted by \includegraphics[height=1.5ex]{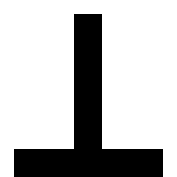} in the figure, live on an infinite number of equally spaced slip planes at $y \in \{0,\pm h,\pm 2h, \ldots,\pm \infty\}$. The dislocations lines are all assumed to be straight and perpendicular to the $x$--$y$ plane. Their Burgers vectors, which are all of equal length $b$, are aligned with the positive $x$-axis. It is furthermore assumed that the dislocations are arranged in infinite vertical walls and that this wall structure is preserved at all times, i.e.\ the dislocations within a single wall only move collectively and uniformly. The horizontal positions of the walls are denoted $x_i$, where $i = 0, 1, 2, \ldots, n$. The first wall is immobilised at $x = x_0 = 0$ and acts as an obstacle for all other walls; $n$ thus denotes the number of mobile dislocation walls.

The system of dislocation walls is embedded in an infinite linear elastic medium which is characterised by its shear modulus $G$ and Poisson's ratio $\nu$. It is subjected to a remote, constant shear stress $-\tau$. As a result of this applied stress and the interaction between the individual walls, the dislocation walls form a pile-up against the immobile wall at $x=0$. We are interested in establishing the equilibrium pile-up configuration, i.e.\ the positions of the walls at rest.

The stress field emitted by a single dislocation wall can be obtained by summing up the classical expressions for a single dislocation due to Volterra for all dislocations within the wall \citep{Hirth1992}. Since only the shear component contributes to the Peach--Koehler force experienced by another dislocation and since all glide planes within the infinite crystal considered are identical, we limit ourselves to the shear stress acting on the glide plane at $y = 0$. At the position of wall $i$, the stress due to the presence of wall $j$ equals:
\begin{equation} \label{eq:stress}
  \sigma_{xy} = \frac{\bar{G}b}{h}
  \dfrac{\frac{\pi}{h}(x_i-x_j)}%
  {\sinh^2\!\left(\frac{\pi}{h}(x_i-x_j)\right)}
\end{equation}
wherein the elastic constant $\bar{G}$ is defined as
\begin{equation} \label{eq:shear}
  \bar{G} = \frac{G}{2(1-\nu)}
\end{equation}

A given wall $i$, where $1 \leq i \leq n$, reaches equilibrium when the stress fields due to all other walls cancel the externally applied stress, i.e.\ when
\begin{equation} \label{eq:equi.walls}
  \sum_{\substack{j=0\\ j \neq i}}^n \:
  \dfrac{\frac{\pi}{h}(x_i-x_j)}%
  {\sinh^2\!\left(\frac{\pi}{h}(x_i-x_j)\right)}
  = \dfrac{\tau h}{\bar{G} b}
\end{equation}
For the immobile wall at $x=0$ to be in equilibrium, a reaction stress equal to $(n+1)\tau$ must be added to this equation \citep{Eshelby1951}. However, since the position of this wall is already known, the resulting equation can be disregarded in what follows.

The equilibrium equations \eqref{eq:equi.walls}, for $1 \leq i \leq n$, have been solved numerically by adopting a linear drag law for the motion of dislocations and solving the instationary problem corresponding to \eqref{eq:equi.walls} using a forward Euler time discretisation. The initial distribution of dislocation walls was taken equidistant and the simulations were continued until the velocity of the walls was negligible.

The equilibrium positions of the dislocation walls obtained from the numerical simulations are reported below in terms of a dislocation wall density profile. The wall density $f$ at position $x_i$ of wall $i$ ($1 \leq i \leq n-1$) is defined as
\begin{equation} \label{eq:dens}
  f(x_i) = \frac{2}{x_{i+1}-x_{i-1}}
\end{equation}

\begin{figure}[htp]
  \centering
  \begin{minipage}[b]{7.5cm}
    \centering
    \includegraphics[width=1.\textwidth]{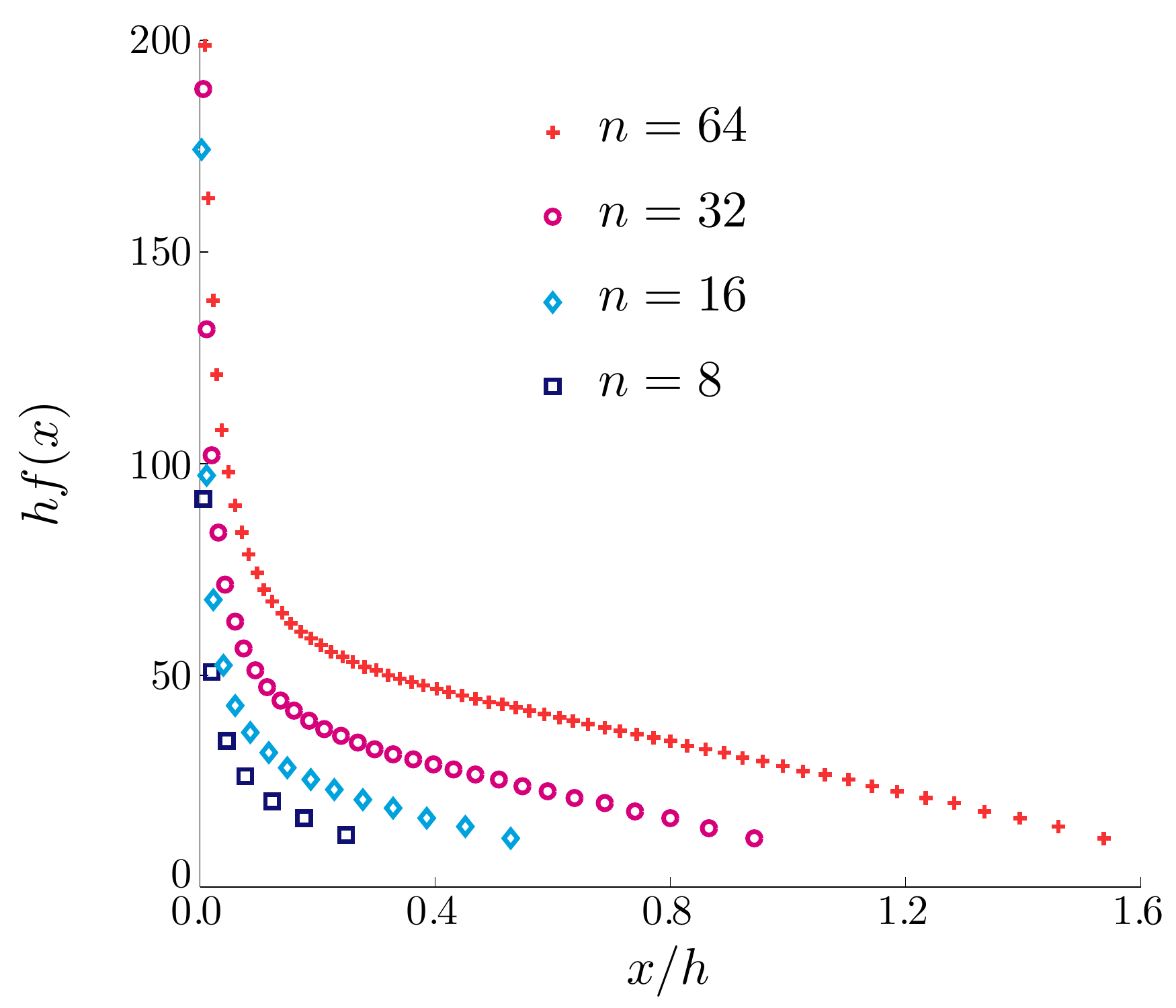}
    \\
    \footnotesize (a) $n$ varied, $(\tau h)/(\bar{G} b) = 8$
  \end{minipage}
  \hfill
  \begin{minipage}[b]{7.5cm}
    \centering
    \includegraphics[width=1.\textwidth]{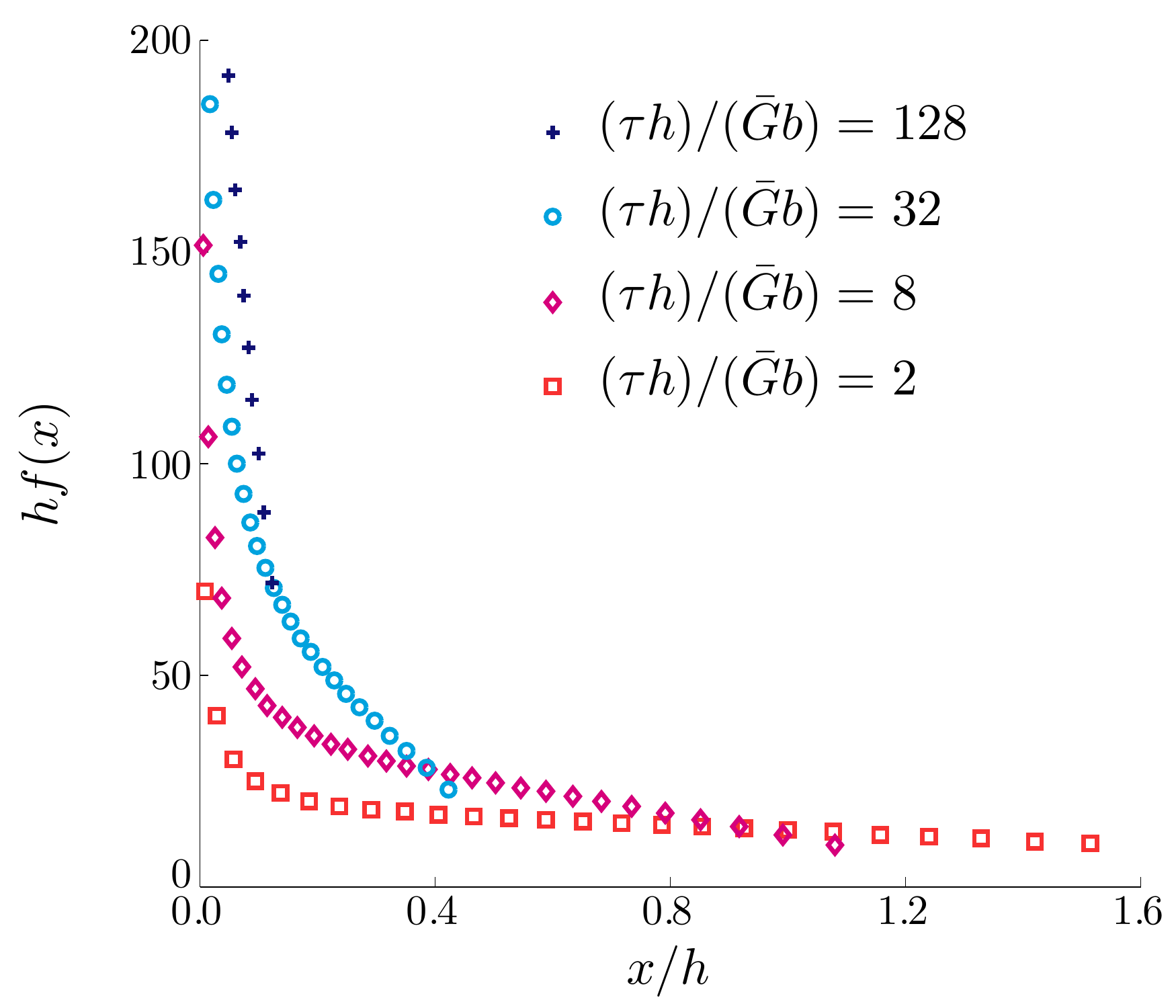}
    \\
    \footnotesize (b) $(\tau h)/(\bar{G} b)$ varied, $n = 32$
  \end{minipage}
  \caption{Numerically obtained dislocation wall densities in the equilibrium   state. The markers indicate the density $f$ at the position $x_i$ of the mobile dislocation wall in equilibrium. Both axes have been normalised by the internal vertical spacing, $h$.}
  \label{fig:num}
\end{figure}

A selection of the simulation results is presented in Fig.~\ref{fig:num}. In these diagrams, the number of mobile dislocations walls $n$ and the normalised stress $(\tau h)/(\bar{G} b)$ are varied, respectively. The position is normalised by the internal vertical spacing $h$, as suggested by \eqref{eq:equi.walls}, and so is the density. A reference solution that appears in both diagrams is based on $n = 32$ mobile dislocation walls and $(\tau h)/(\bar{G} b) = 8$.

In Fig.~\ref{fig:num}(a), the number of mobile dislocation walls $n$ is varied relative to the reference solution; the other parameters of the problem are constant. We observe that for larger $n$ the length of the pile-up is larger, while the slope of the density profile tail appears to remain identical.

Fig.~\ref{fig:num}(b) shows the combined influence of the internal vertical spacing $h$ of the walls and the applied shear stress $\tau$, normalised by the shear modulus $\bar{G}$ and the Burger's vector $b$. A larger ratio $(\tau h)/(\bar{G} b)$ causes the pile-up to be confined to a smaller domain, while the slope of the density is larger.

In both diagrams, the dislocation wall density profile exhibits two distinct regions. At some distance from the obstacle, the density $f(x)$ appears to decay linearly. However, close to the obstacle $f(x)$ shows a much stronger variation and it appears to be singular at $x = 0$. Below we study each of these two regimes more closely and obtain analytical expressions for them.

\section{Analysis of the region near the obstacle}
\label{sec:near}

We first focus on the region near the obstacle, where the computed wall densities of Fig.~\ref{fig:num} appear to show a singularity. The behaviour in this regime can be understood by realising that near the obstacle the horizontal spacing between the individual walls is much smaller than the vertical spacing $h$. As a consequence, the interaction between different dislocation walls takes place predominantly along the individual glide planes. Thus stresses due to dislocations on other glide planes may be neglected. This observation allows us to resort to the classical analysis of a pile-up on a single glide plane.

\begin{figure}[htp]
  \centering
  \begin{minipage}[b]{7.5cm}
    \centering
    \includegraphics[width=1.\textwidth]{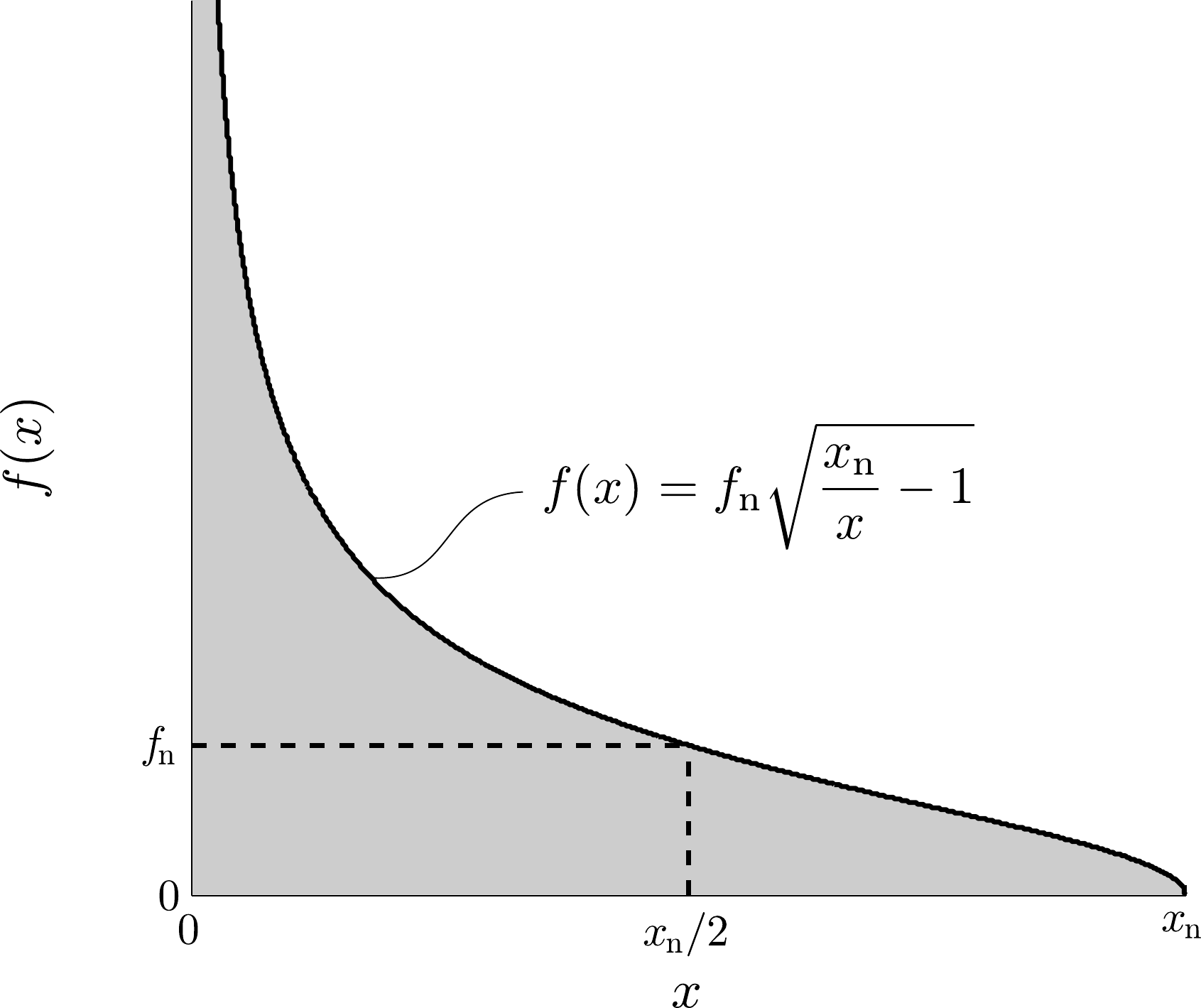}
    \\
    \footnotesize (a) near region
  \end{minipage}
  \hfill
  \begin{minipage}[b]{7.5cm}
    \centering
    \includegraphics[width=1.\textwidth]{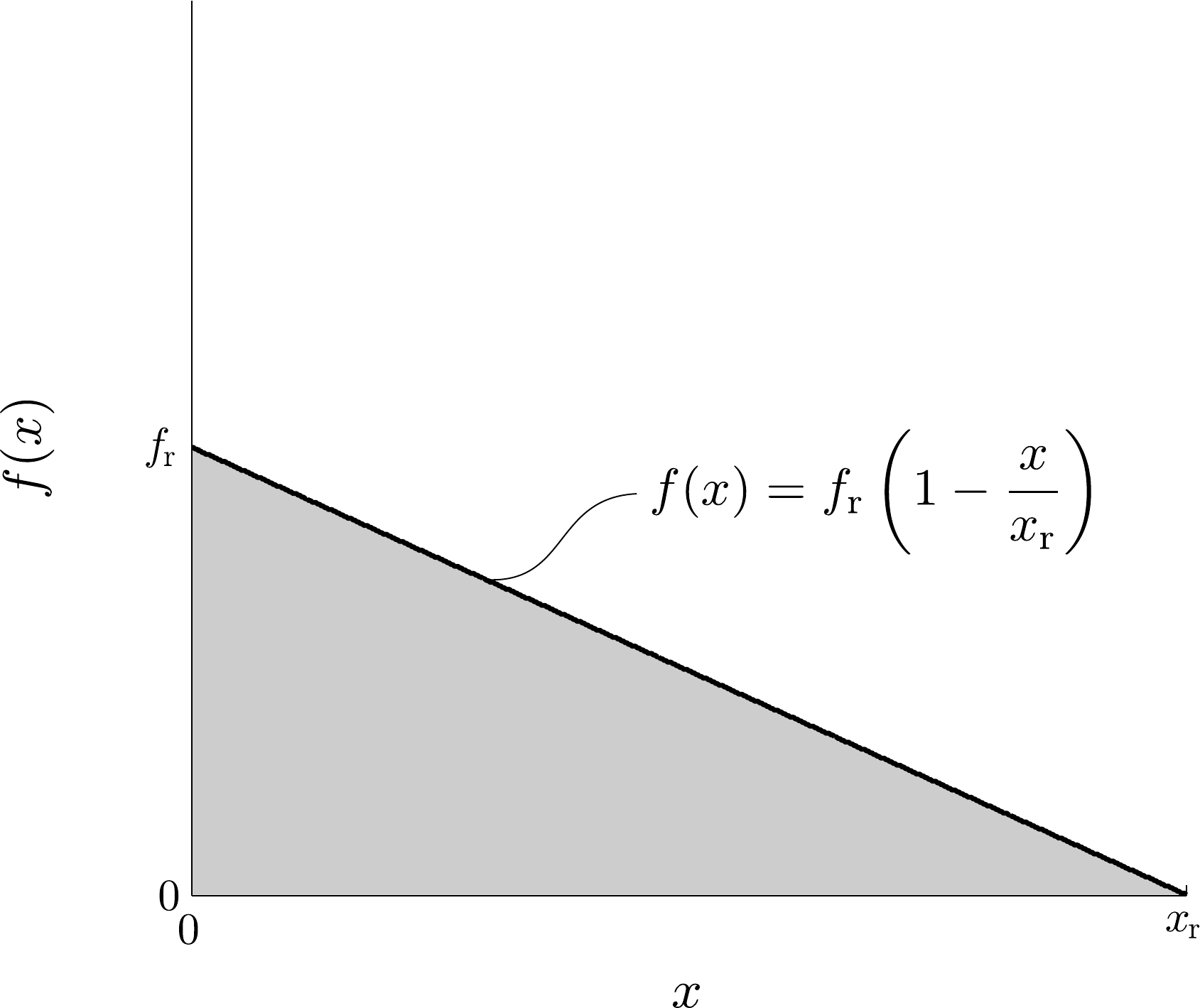}
    \\
    \footnotesize (b) remote region
  \end{minipage}
  \caption{Sketch of the dislocation wall density profiles established for the two regions distinguished in the pile-up.}
  \label{fig:dens_reg}
\end{figure}

In terms of the parameters used here, the classical continuous solution for a single glide plane reads \citep{Leibfried1951,Head1955}:
\begin{equation} \label{eq:sin}
  f(x) = f_\mathrm{n} \, \sqrt{\dfrac{x_\mathrm{n}}{x} - 1}
\end{equation}
where
\begin{equation} \label{eq:sin.len}
  x_\mathrm{n} = \dfrac{ 2 n \bar{G} b}{\tau \pi}
  \qquad
  f_\mathrm{n} = \dfrac{\tau}{\bar{G} b}
\end{equation}
A sketch of the above dependence is shown in Fig.~\ref{fig:dens_reg}(a). The constant $x_\mathrm{n}$ represents the end of the (single glide plane) pile-up, at which $f = 0$.

In our numerical results, a transition to a different (linear) regime is generally observed well before the density reaches zero. We can therefore assume $x \ll x_\mathrm{n}$ and simplify \eqref{eq:sin} to
\begin{equation} \label{eq:sin.approx}
  f(x) \approx f_\mathrm{n} \, \sqrt{\dfrac{x_\mathrm{n}}{x}}
\end{equation}
The behaviour of $f(x)$ near the obstacle is thus expected to exhibit a singularity proportional to $1/\sqrt{x}$.

\begin{figure}[htp]
  \centering
  \begin{minipage}[b]{7.5cm}
    \centering
    \includegraphics[width=1.\textwidth]{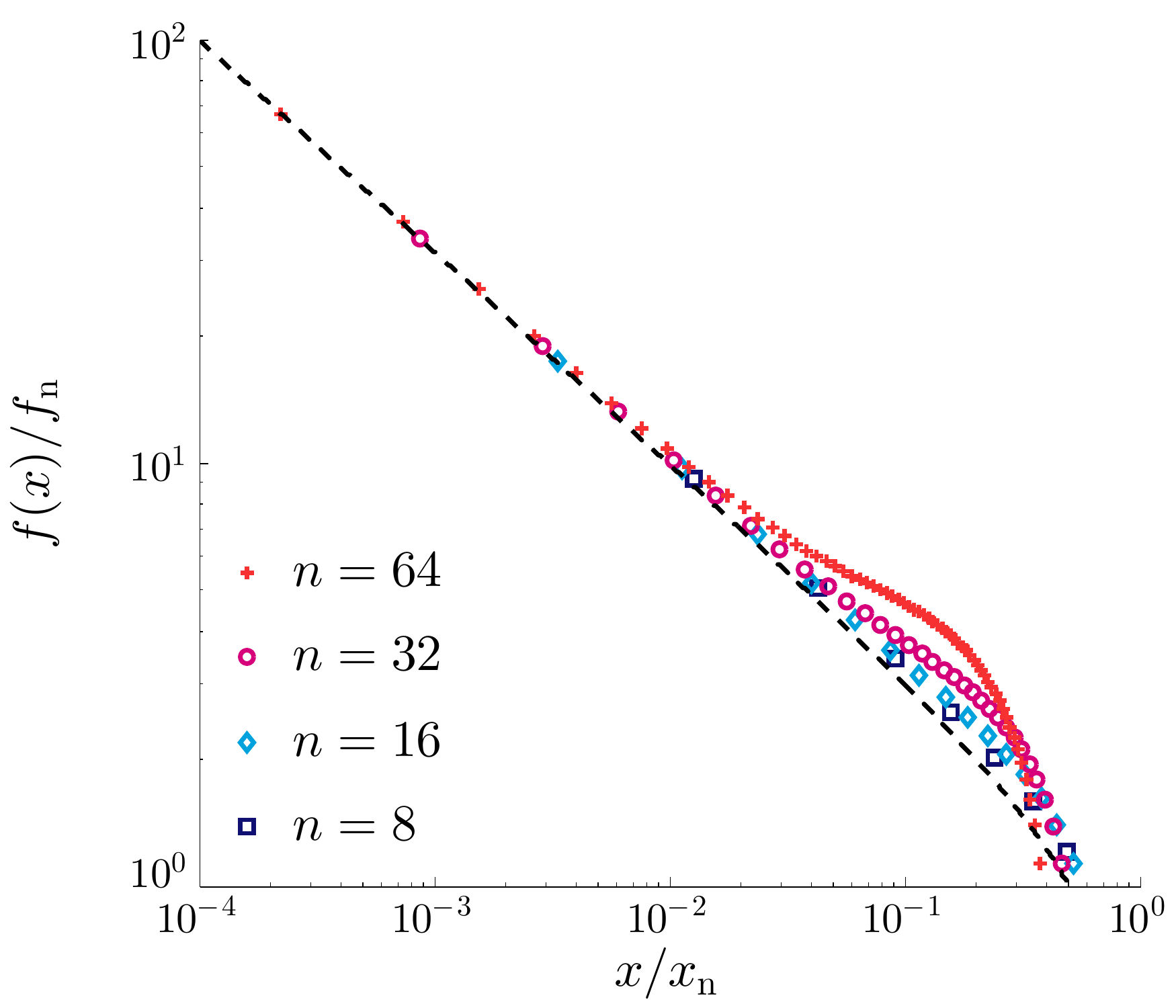}
    \\
    \footnotesize (a) $n$ varied, $(\tau h)/(\bar{G} b) = 8$
  \end{minipage}
  \hfill
  \begin{minipage}[b]{7.5cm}
    \centering
    \includegraphics[width=1.\textwidth]{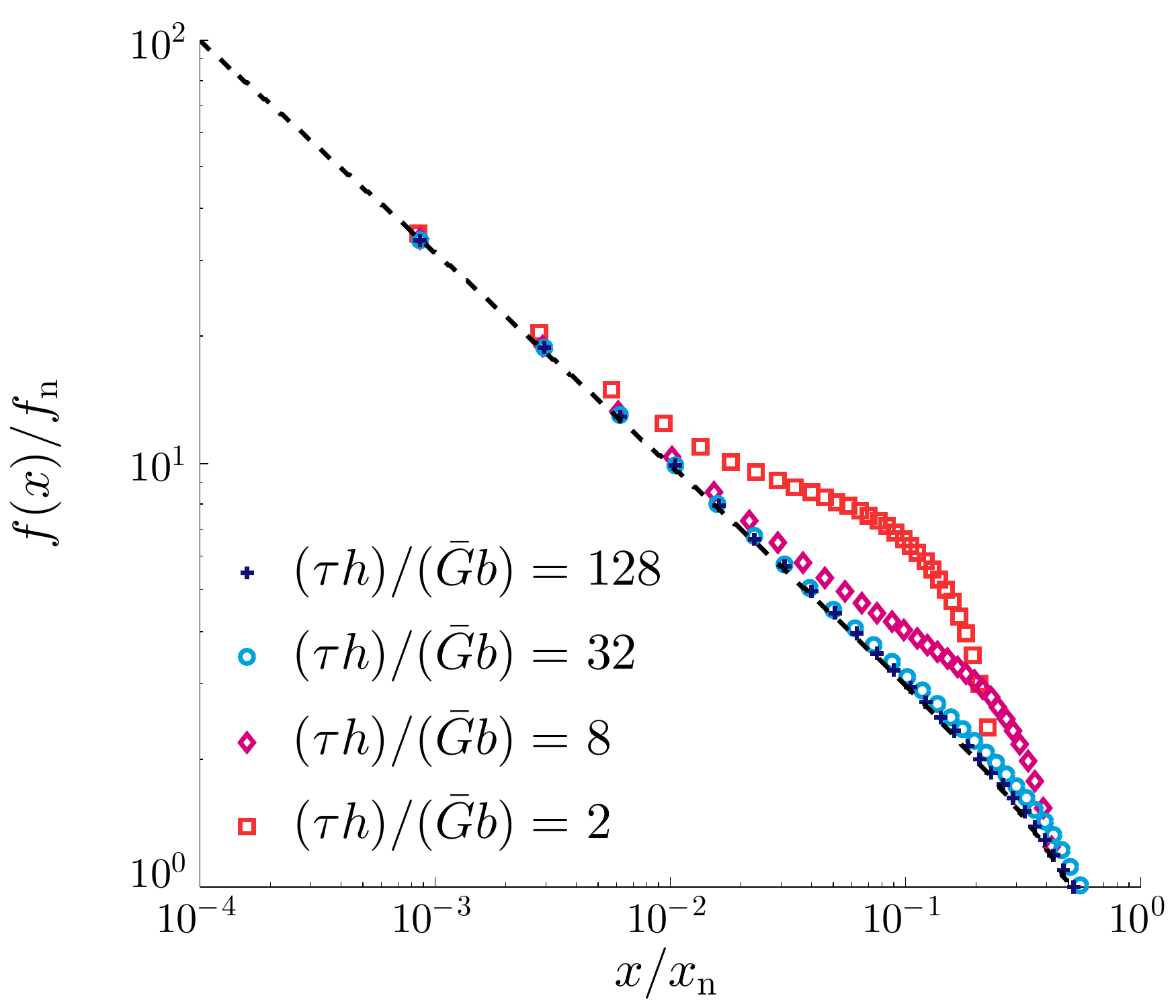}
    \\
    \footnotesize (b) $(\tau h)/(\bar{G} b)$ varied, $n = 32$
  \end{minipage}
  \caption{The numerical data scaled by $x_\mathrm{n}$ and $f_\mathrm{n}$ as derived from the analysis of the near pile-up region. The dashed curve represents the analytical solution of a single glide plane \eqref{eq:sin}.}
  \label{fig:near}
\end{figure}

Fig.~\ref{fig:near} shows the numerical data (of Fig.~\ref{fig:num}), normalised by $x_\mathrm{n}$ and $f_\mathrm{n}$ as given by \eqref{eq:sin.len}; note the logarithmic axes used in the diagram. The analytical solution for a single glide plane, given in \eqref{eq:sin}, is shown as a dashed curve. For small $x$, i.e.\ near the obstacle, the numerical data follows this curve well for virtually all parameter values considered. For most simulations, however, the fraction of data points in this singular region is limited, and the data starts to deviate from the prediction at some distance from the obstacle as it enters the remote region (see the next section).

The best correspondence, i.e.\ the largest singular region, is observed for a large vertical spacing $h$ and a large applied stress $\tau$ (Fig.~\ref{fig:near}(b)) -- see also Fig.~\ref{fig:num}. In the former case the criterion that the horizontal spacing of dislocations is smaller than their vertical spacing is met more easily. In the latter case, in the limit of high applied stress, the dislocation walls are squeezed towards the obstacle and their horizontal distance is therefore small. Note that this is the regime considered predominantly by \citet{Roy2008}, who indeed found a $1/\sqrt{x}$ distribution.

The length scale governing the near part of the pile-up is given by the quotient $n \bar{G} b / \tau$ -- cf.\ expression (\ref{eq:sin.len}a) for the length of the pile-up. The solution is independent of the slip plane spacing $h$ since interactions between the different slip planes are negligible.

\section{Analysis of the region remote to the obstacle}
\label{sec:remote}

We now consider the region remote to the obstacle, where a more or less linear wall density profile is observed in the numerical results of Fig.~\ref{fig:num}. We demonstrate below that sufficiently far away from the two ends of the pile-up, a linear decay according to
\begin{equation} \label{eq:fr}
  f(x) = f_\mathrm{r} \left( 1 - \dfrac{x}{x_\mathrm{r}} \right)
\end{equation}
indeed is a good approximation of the equilibrium solution. In this equation the constant $x_\mathrm{r}$ represents the length of the pile-up and $f_\mathrm{r}$ is a constant. Note that the same, linear density profile was established by \citet{Hall2011a}.

To verify this approximation, we assume \eqref{eq:fr} to describe entire domain of the pile-up. This implies that in order to represent a total of $n$ walls within the pile-up domain $(0, x_\mathrm{r})$, we need to satisfy
\begin{equation} \label{eq:lin.condition}
  \int\limits_0^{x_\mathrm{r}} \! f(x) \: \mathrm{d}x =
  \frac{f_\mathrm{r} \, x_\mathrm{r}}{2} = n
\end{equation}
(see also Fig.~\ref{fig:dens_reg}(b)). By making this assumption we fail to take the singular part of the pile-up, near the obstacle, into account. This introduces a minor approximation as a small fraction of dislocation walls are located in this region. Notice how this assumption was implicitly also made in the previous section where \eqref{eq:sin} was assumed to describe the entire domain of the pile-up. The remaining unknown, the pile-up length $x_\mathrm{r}$, follows from our analysis below.

To be able to use the continuous density function $f(x)$ defined in \eqref{eq:fr}, we first rewrite the discrete equilibrium problem \eqref{eq:equi.walls} in the following continuous form:
\begin{equation} \label{eq:equi.walls.con}
  \int\limits_{0}^{x_\mathrm{r}} \! f(\tilde{x}) \,
  \dfrac{\frac{\pi}{h}(x-\tilde{x})}%
  {\sinh^2\!\left(\frac{\pi}{h}(x-\tilde{x})\right)}
  \: \mathrm{d}\tilde{x}
  = \dfrac{\tau h}{\bar{G} b}
\end{equation}
Indeed, upon insertion of the density $f(\tilde{x}) ~ \text{d}\tilde{x} = \sum_j \delta(\tilde{x} - x_j)$ this equation reduces to the original discrete form \eqref{eq:equi.walls}.

It is now convenient to introduce the dimensionless relative distance
\begin{equation} \label{eq:relpos}
  \xi = \dfrac{\pi}{h} \, (\tilde{x} - x)
\end{equation}
and the dimensionless density
\begin{equation}
  \phi(\xi) = h \, f\!\left(x + \frac{h}{\pi} \xi\right)
\end{equation}
With these definitions we rewrite \eqref{eq:equi.walls.con} as
\begin{equation} \label{eq:equi.walls.con.inter}
  I(x) := -\dfrac{3}{\pi^2}
  \int\limits_{-\xi_0}^{\xi_\mathrm{r}} \! \phi(\xi) \,
  \dfrac{\xi}{\sinh^2\!\xi} \: \mathrm{d}\xi = \dfrac{3\tau h}{\pi \bar{G} b}
\end{equation}
where the limits of the integral are given by
\begin{equation}
  \xi_0 = \dfrac{\pi}{h} \, x
  \qquad
  \xi_\mathrm{r} = \dfrac{\pi}{h} \left(x_\mathrm{r} - x\right)
\end{equation}

\begin{figure}[htp]
  \centering
  \includegraphics[width=8cm]{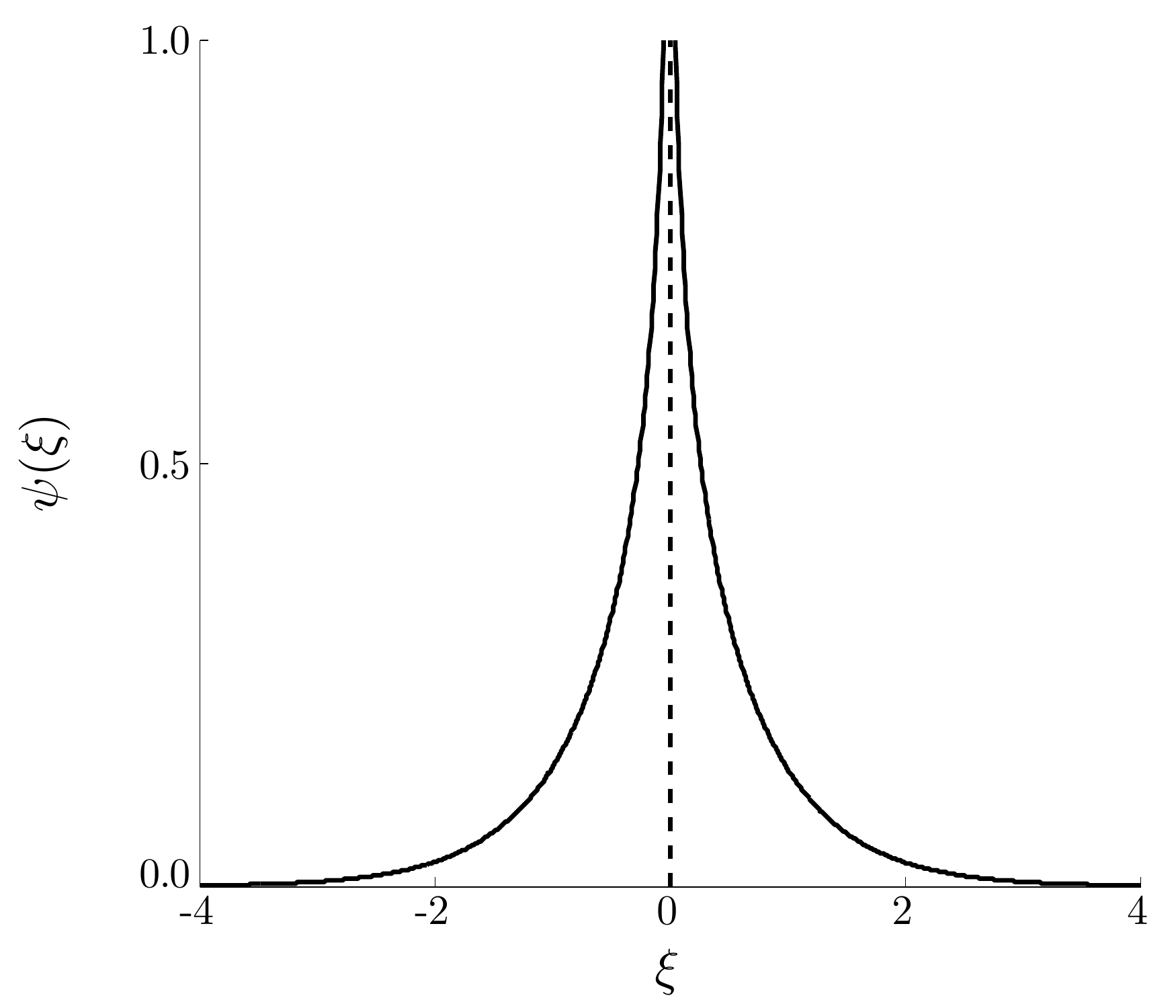}
  \caption{Auxiliary function $\psi(\xi)$.}
  \label{fig:psi}
\end{figure}

The factor $3/\pi^2$ entails a convenient property that becomes apparent by introducing the following auxiliary function
\begin{equation} \label{eq:psi.def}
  \psi(\xi) =
  \frac{3}{\pi^2} \biggl( \xi \coth\xi - \log\!\left|2\sinh\xi\right| \biggr)
\end{equation}
which is sketched in Fig.~\ref{fig:psi}. This even function satisfies
\begin{equation} \label{eq:psi.prop}
  \psi^\prime(\xi) = -\dfrac{3}{\pi^2} \dfrac{\xi}{\sinh^2\xi}
  \qquad
  \lim\limits_{|\xi| \rightarrow \infty} \psi(\xi) = 0
  \qquad
  \int\limits_{-\infty}^{\infty} \! \psi(\xi) \: \mathrm{d}\xi = 1
\end{equation}
The latter property, (\ref{eq:psi.prop}c), stems from the factor $3/\pi^2$; a proof is given in the appendix.

Using this auxiliary function, the integral \eqref{eq:equi.walls.con.inter} further reduces to
\begin{equation}
  I(x) = \int\limits_{-\xi_0}^{\xi_\mathrm{r}} \! \phi(\xi) \: \psi^\prime(\xi)
  \: \mathrm{d}\xi
\end{equation}
or, integrating by parts,
\begin{equation} \label{eq:int.part}
  I(x) = \phi(\xi) \, \psi(\xi) \biggr|_{-\xi_0}^{\xi_\mathrm{r}}
  - \int\limits_{-\xi_0}^{\xi_\mathrm{r}} \! \phi^\prime(\xi) \,
  \psi(\xi) \: \mathrm{d}\xi
\end{equation}

Sufficiently far away from the head and tail of the pile-up $\xi_0$ and $\xi_\mathrm{r}$ are large. A reasonable approximation is therefore to take the limit $\xi_0, \xi_\mathrm{r} \rightarrow \infty$; we will discuss the implications of this approximation in the next section. Consequently, property (\ref{eq:psi.prop}b) causes the first term in \eqref{eq:int.part} to vanish. Furthermore, as the derivative $\phi^\prime(\xi)$ of the linear density profile is constant and \eqref{eq:lin.condition} and (\ref{eq:psi.prop}c) hold, we arrive at the approximation
\begin{equation} \label{eq:int.sol}
  I(x) \approx - \phi^\prime(\xi) \int\limits_{-\infty}^{\infty} \!
  \psi(\xi) \: \mathrm{d}\xi =
  -\phi^\prime(\xi) = \dfrac{2nh^2}{\pi x_\mathrm{r}^2}
\end{equation}

Inserting the approximation \eqref{eq:int.sol} in the equilibrium equation \eqref{eq:equi.walls.con.inter} finally allows us to determine the constants $x_\mathrm{r}$ and, again via \eqref{eq:lin.condition}, $f_\mathrm{r}$ in terms of the parameters of the problem:
\begin{equation} \label{eq:lin.len}
  x_\mathrm{r} = \sqrt{\dfrac{2 n \bar{G} b h}{3 \tau}}
  \qquad
  f_\mathrm{r} = \sqrt{\dfrac{6 n \tau}{\bar{G} b h}}
\end{equation}
It can easily be verified that these constants are consistent with those derived in \citep{Hall2011a}.

Contrary to the near region, the quotient $n \bar{G} b / \tau$ does not represent the only relevant length scale in the remote region. The length of the pile-up is now set by the square root of its product with the slip plane spacing $h$ (cf.\ (\ref{eq:lin.len}a)). The slope of the wall density distribution is predicted to be
\begin{equation}
  \dfrac{f_\mathrm{r}}{x_\mathrm{r}} = \dfrac{3\tau}{\bar{G}bh}
\end{equation}
In particular, it is steeper for a larger applied stress $\tau$ or smaller vertical spacing $h$ and it does not depend on the number of walls $n$. This corresponds well with the trends observed in Fig.~\ref{fig:num}.

\begin{figure}[htp]
  \centering
  \begin{minipage}[b]{7.5cm}
    \centering
    \includegraphics[width=1.\textwidth]{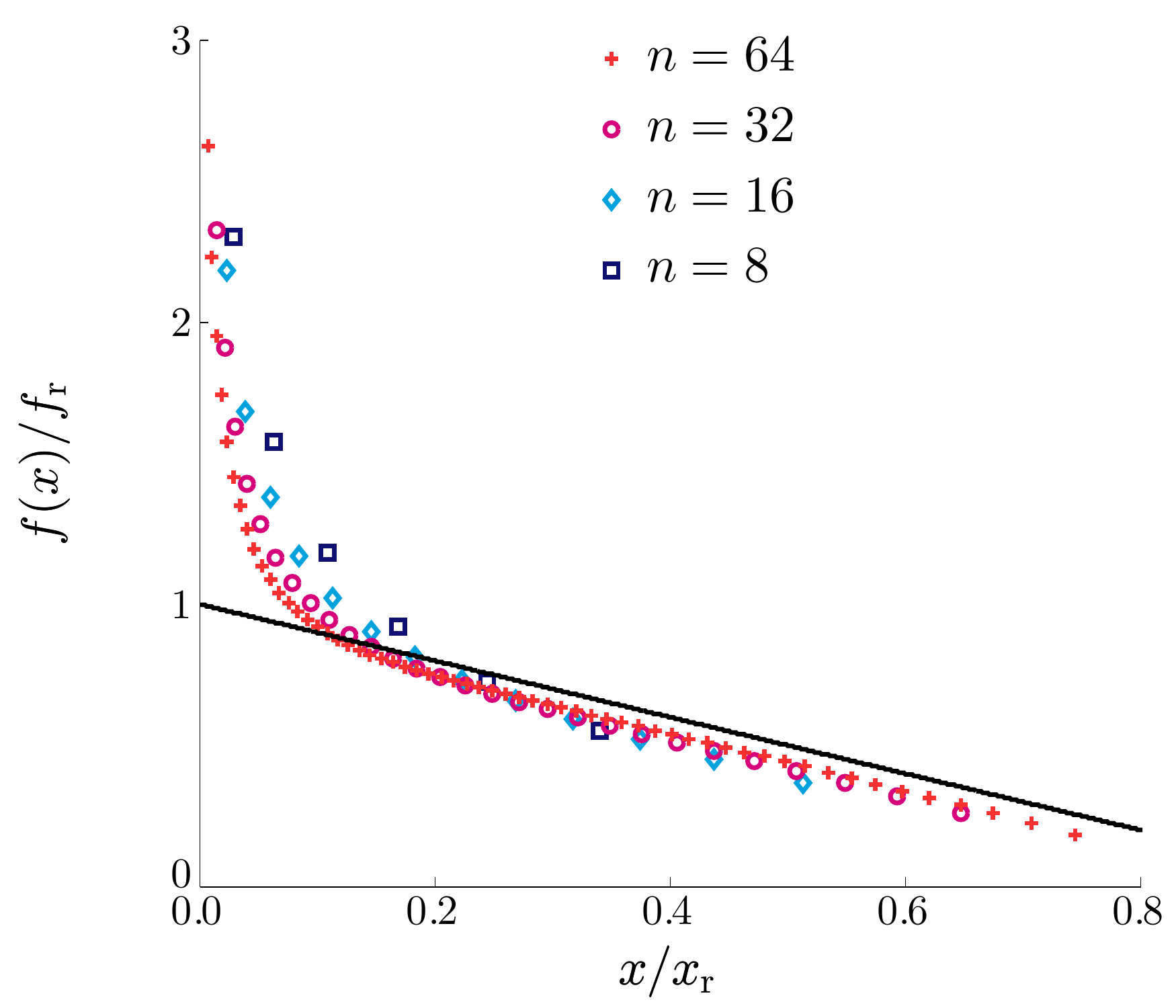}
    \\
    \footnotesize (a) $n$ varied, $(\tau h)/(\bar{G} b) = 8$
  \end{minipage}
  \hfill
  \begin{minipage}[b]{7.5cm}
    \centering
    \includegraphics[width=1.\textwidth]{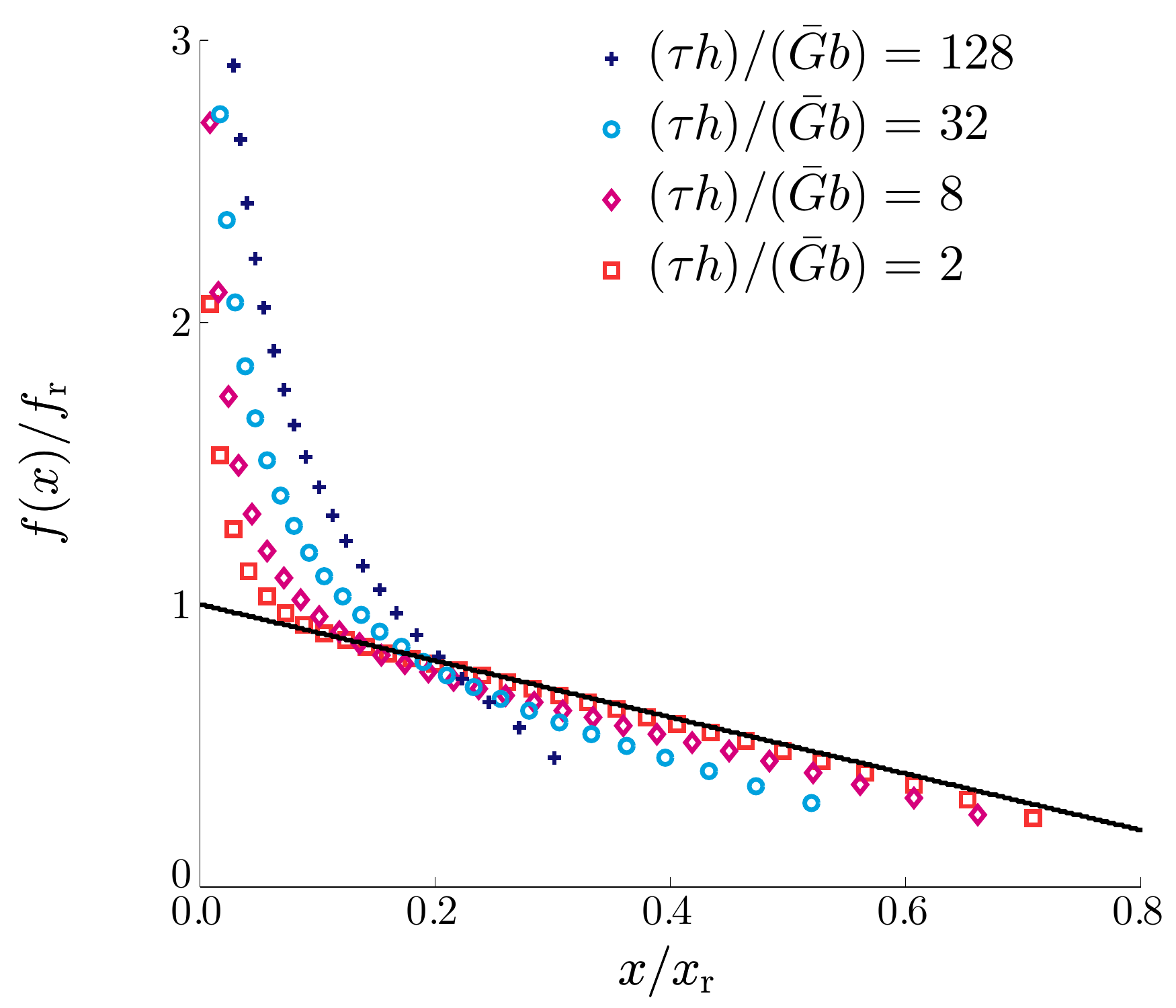}
    \\
    \footnotesize (b) $(\tau h)/(\bar{G} b)$ varied, $n = 32$
  \end{minipage}
  \caption{The numerical results, normalised according to the scaling suggested by the analysis of the remote region. The black solid line represents the predicted density distribution of \eqref{eq:fr}.}
  \label{fig:remote}
\end{figure}

In order to validate the approximation $f(x)$ of \eqref{eq:fr}, the numerical data of Fig.~\ref{fig:num} has been scaled by the respective constants $x_\mathrm{r}$ and $f_\mathrm{r}$ as given by \eqref{eq:lin.len}; the result is shown in Fig.~\ref{fig:remote}. The remote parts of most of the normalised numerical pile-up data in these diagrams indeed collapse onto the single straight line predicted by \eqref{eq:fr}. Some deviation is observed for those solutions in which most of the dislocation walls occupy the near regime, see e.g.\ the data for $(\tau h)/(\bar{G} b) = 128$ in Fig.~\ref{fig:remote}(b). Under these conditions relation~\eqref{eq:lin.condition} becomes inaccurate and our analysis thus breaks down -- see also the next section.

\section{Transition region and combined prediction}
\label{sec:transition}

\begin{figure}[htp]
  \centering
  \begin{minipage}[b]{7.5cm}
    \centering
    \includegraphics[width=1.\textwidth]{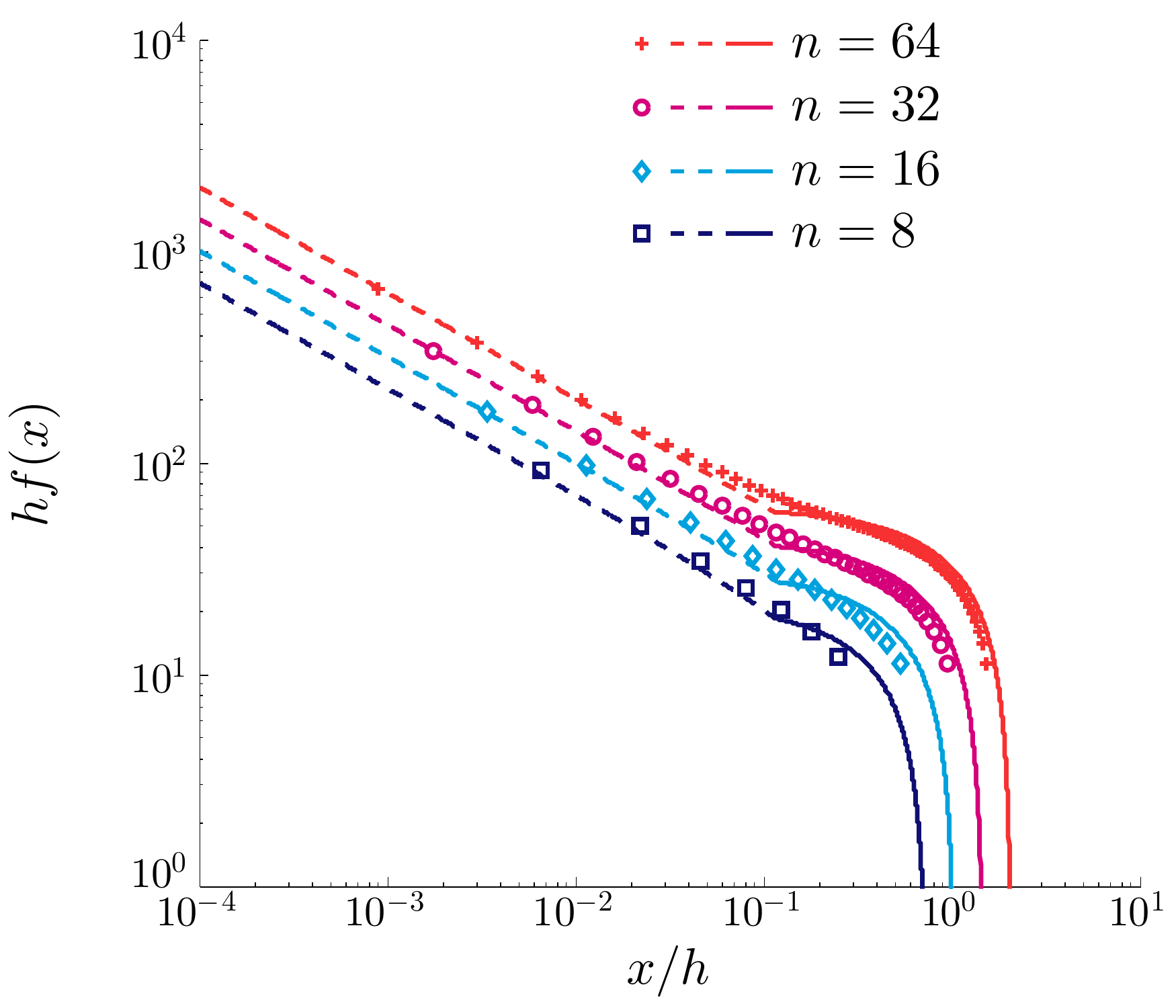}
    \\
    \footnotesize (a) $n$ varied, $(\tau h)/(\bar{G} b) = 8$
  \end{minipage}
  \hfill
  \begin{minipage}[b]{7.5cm}
    \centering
    \includegraphics[width=1.\textwidth]{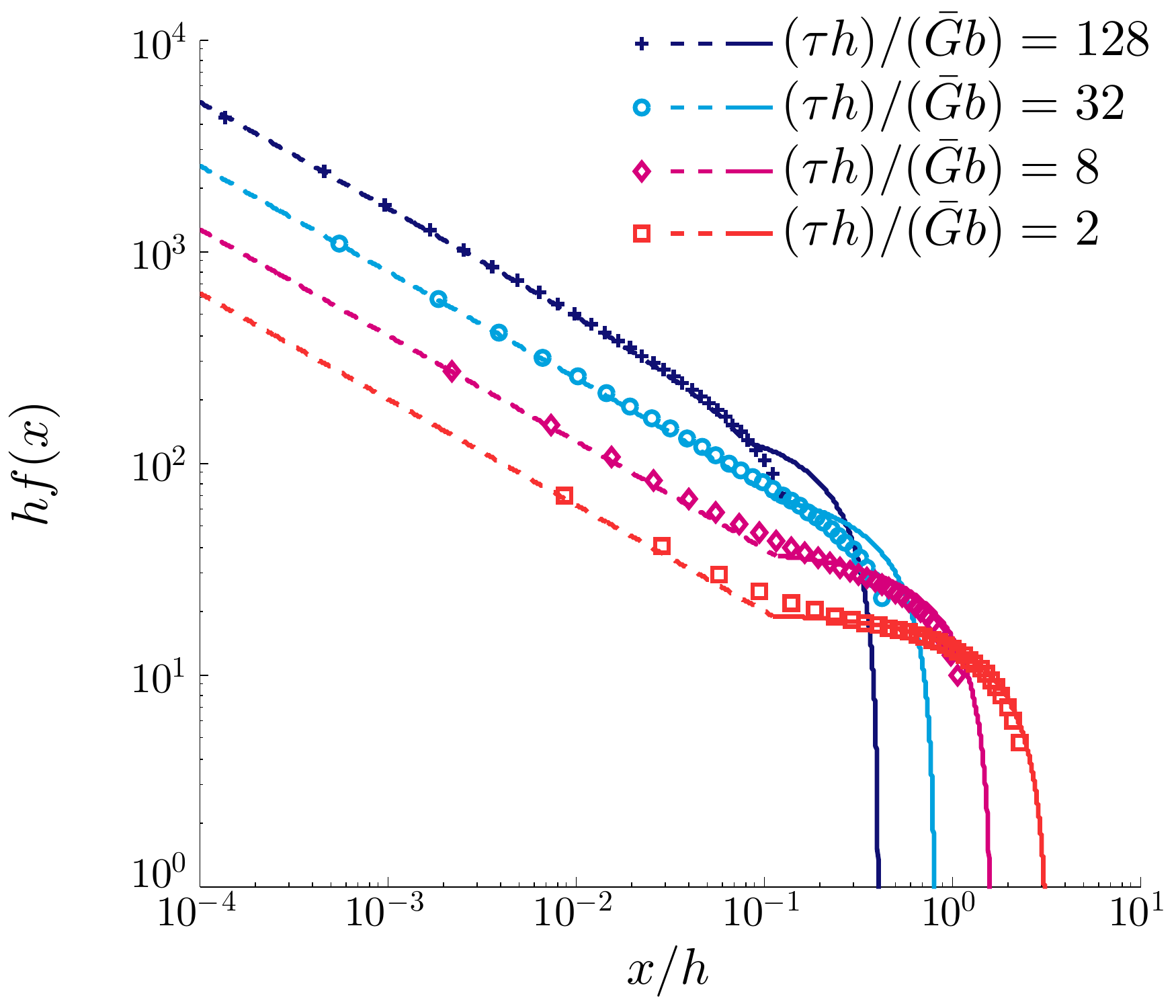}
    \\
    \footnotesize (b) $(\tau h)/(\bar{G} b)$ varied, $n = 32$
  \end{minipage}
  \caption{Analytical solutions of the near (dashed curves) and remote regions (solid curves) compared with the discrete numerical data.}
  \label{fig:trans}
\end{figure}

So far, we have established separate closed-form expressions for the dislocation wall density near and remote to the obstacle, given by Equations~\eqref{eq:sin}, \eqref{eq:sin.len} and \eqref{eq:fr}, \eqref{eq:lin.len} respectively. For the same parameter sets used in the numerical analysis of Fig.~\ref{fig:num}, Fig.~\ref{fig:trans} shows the predicted density profiles normalised by the internal vertical spacing $h$ of the walls, on the relevant part of the domain, superimposed on the corresponding numerical data. With both the remote and the near region predicted quite accurately by the analytical expressions, we now seek to understand transition between these two regions.

An estimate of the distance from the obstacle at which the transition between the singular and linear profiles takes place may be obtained by revisiting the analysis of the remote region. As a major assumption, we approximated the limits $-\xi_0$ and $\xi_\mathrm{r}$ in \eqref{eq:int.part} by minus and plus infinity respectively. This approximation is justified when $\xi_0$ and $\xi_\mathrm{r}$ are sufficiently large, since in this case the left and right tail of $\psi(\xi)$ contribute negligibly to the integral. Near the obstacle, $\xi_\mathrm{r}$ may still be sufficiently large, but replacing $\xi_0$ by infinity becomes increasingly questionable since the added (left) tail may contribute significantly. Quantitatively, when we replace $\xi_0 = 1$ by $\xi_0 = \infty$ an error of less than 10\% is introduced in evaluating the integral of $\psi(x)$. A more significant error of 25\% is introduced at $\xi_0 \approx \tfrac{1}{3}$.

In terms of the unscaled coordinate $x$, the above implies that the distance between the obstacle and the transition point scales with $h$ and that a significant deviation from the predicted linear trend in the remote region may be expected at distances smaller than $x = \tfrac{1}{3}h/\pi \approx h/10$. This is consistent with the data of Fig.~\ref{fig:trans}, in which the normalised horizontal position $x/h$ of the transition is virtually constant for variations in $n$ and $(\tau h)/(\bar{G} b)$. In both diagrams we observe that the transition takes place at approximately $x/h = 1/10$.

A more physical interpretation of the transition between the two regimes can be given as follows. Sufficiently far away from the obstacle, a dislocation wall interacts with other dislocation walls within a band on the order of $h/10$ to the left and to the right. Beyond this band, the mutual interactions become too weak to contribute significantly to the Peach-Koehler force experienced by the wall. In our pile-up problem, the repelling forces exerted on a certain wall by walls on its left must exceed those from the right, so that the nett force balances the applied stress $-\tau$. In order to generate this positive nett interaction force, the system develops a negative gradient in dislocation density. As a result of this gradient, the distance between dislocation walls on the left of a certain wall is smaller than that distance on the right. This implies that (i) the individual interactions are stronger and (ii) more walls fit in the interaction band on the left compared to the right and thus generates the necessary nett force.

Near the obstacle, the interaction band extends beyond the obstacle, where no dislocation walls exist. This implies that the force from the left must be delivered by fewer dislocation walls -- in fact by a number which decreases as we approach the obstacle. To nevertheless generate the force required, these walls need to have a considerably smaller horizontal spacing, so that the force per wall is higher (cf.\ \eqref{eq:stress}). This explains why the wall density increases so dramatically towards the obstacle.

\section{Summary and concluding remarks}
\label{sec:conclusion}

In this paper we have analysed the pile-up of infinite walls of edge dislocations against an obstacle which is parallel to the walls and perpendicular to the slip planes. Numerical solutions of the discrete equilibrium problem, similar to those of \citet{Roy2008}, show two distinct regions. In the direct vicinity of the obstacle, within a bandwidth which scales with the (active) slip plane spacing $h$, a $1/\sqrt{x}$ decay of the dislocation wall density is observed. This part corresponds with the classical case of a linear pile-up on a single slip plane \citep{Eshelby1951,Leibfried1951,Head1955}. Outside this band, remote to the obstacle, a linear distribution is obtained \citep{Hall2011a}. Combining the partial solutions obtained for the two regions allows one to predict the wall density profile of the entire pile-up with quite reasonable accuracy, as evidenced by a comparison with the discrete data (in Fig.~\ref{fig:trans}).

Our study raises the question which length scale sets the density distribution. In the near-obstacle region, the density decays over a distance which scales with $n \bar{G} b / \tau$. The slip plane spacing $h$ does not play a role in this region, since the parallel pile-ups on the individual slip planes are independent. Beyond this thin boundary layer, however, the density distribution is governed by a different length scale and the length of the pile-up scales with $\sqrt{n \bar{G} b / \tau}$ and with $\sqrt{h}$.

Our results suggest that sufficiently far away from the obstacle, each dislocation wall shows significant interaction with neighbouring walls within a band of approximately $h/10$ to the left and to the right, where $h$ is the vertical spacing of the dislocations within the wall. Beyond this distance, the wall's stress field has decayed sufficiently for it to be neglected. For dislocation walls near the obstacle, the interaction band interferes with the obstacle and a different behaviour is thus observed. At the same time, the horizontal distance between individual dislocations has become smaller than their vertical spacing and interactions between the different slip planes are small compared to those on the same slip plane. As a result, a transition to the classical solution for a single glide plane occurs.

The notion of an interaction band of a thickness which scales with $h$ is consistent with the observation made by \citet{Roy2008} that taking into account only interactions between neighbouring walls results substantially different density profiles. Indeed, the numerical simulations done by \citet{Roy2008} based on a nearest-neighbour assumption have been repeated for the parameter sets used here, yielding similar results (not shown here). The pile-up length obtained in them is significantly smaller than that of the case of full interaction. Furthermore, the density profile obtained matches well with the exponential decay predicted by statistical mechanics arguments \citep{Roy2008,Groma2003} -- unlike the profiles obtained in our full-interaction study. Whether this discrepancy is due to the highly idealised nature of the problem considered here, or has a more fundamental basis, is a matter for further study.

\section*{Acknowledgements}

The authors are grateful to L.~Scardia and  M.A.~Peletier for stimulating and
useful discussions.

\appendix

\section{Normalisation of the auxiliary function $\psi(\xi)$}
\label{appendix/relations}

In this appendix we show that the integral
\begin{equation}
  J := \int\limits_{-\infty}^{\infty} \! \psi(\xi) \: \mathrm{d}\xi
\end{equation}
with $\psi(\xi)$ given by \eqref{eq:psi.def}, equals one, as stated in Equation~(\ref{eq:psi.prop}c).

The evenness of $\psi(\xi)$ allows us to rewrite $J$ as
\begin{equation}
  J = 2 \int\limits_{0}^{\infty} \! \psi(\xi) \: \mathrm{d}\xi =
  \dfrac{6}{\pi^2} \int\limits_{0}^{\infty} \biggl[ \xi \coth\xi
   - \log(2\sinh\xi) \biggr] \: \mathrm{d}\xi
\end{equation}
or, using the substitution $2\xi = z$,
\begin{equation} \label{eq:g.sub}
  J = \dfrac{3}{\pi^2} \int\limits_{0}^{\infty}
  \left[ \dfrac{z e^{-z}}{1 - e^{-z}} - \log\left(1 - e^{-z}\right) \right]
  \mathrm{d}z
\end{equation}

Now use the fact that
\begin{equation}
  \dfrac{\mathrm{d}}{\mathrm{d}z}
  \biggl[ z \log\left(1 - e^{-z}\right) \biggr] =
  \dfrac{z e^{-z}}{1-e^{-z}} + \log\left(1 - e^{-z}\right)
\end{equation}
to eliminate the first term in \eqref{eq:g.sub}. This results in
\begin{equation}
  J = \dfrac{3}{\pi^2} \, z \, \log\left( 1 - e^{-z} \right) \biggr|_0^{\infty}
  - \dfrac{6}{\pi^2} \int\limits_{0}^{\infty} \log\left(1 - e^{-z}\right) \:
  \mathrm{d}z
\end{equation}
The first term in this expression vanishes and the integral in the second equals \citep{Gradshteyn1965}
\begin{equation}
  \int\limits_0^\infty \log\left(1 - e^{-z}\right) \: \mathrm{d}z =
  -\frac{\pi^2}{6}
\end{equation}
so that we finally have $J = 1$.

\bibliography{library}

\end{document}